\documentclass[aps]{revtex4}
\usepackage{amssymb,amsmath,epsfig}
\usepackage[colorlinks=true, pdfstartview=FitV, linkcolor=blue, citecolor=red, urlcolor=magenta]{hyperref}
\usepackage{graphicx}
\usepackage{latexsym}
\usepackage{amsmath}
\usepackage{amsfonts}
\usepackage{amssymb}
\usepackage{verbatim}

\usepackage[all]{xy}

\usepackage{epstopdf}

\newcommand{\be}{\begin{equation}}
\newcommand{\ee}{\end{equation}}
\newcommand{\bea}{\begin{eqnarray}}
\newcommand{\eea}{\end{eqnarray}}

\newcommand{\pa}{\partial}

\newcommand{\bb}{\bibitem}

\def\bb{\bibitem}

\def\ni{\noindent}

\def\bb{\bibitem}
\newcommand{\ben}{\begin{eqnarray}}
\newcommand{\een}{\end{eqnarray}}

\begin{document}
\title{Lifshitz-scaling to Lorentz-violating high derivative operator and gamma-ray busts }

\author{$^{1,3}$E. Passos}
\email{passos@df.ufcg.edu.br}
\author{ $^{4,5}$E. M. C. Abreu}
\email{evertonabreu@ufrrj.br}
\author{$^{1}$M. A. Anacleto}
\email{anacleto@df.ufcg.edu.br}
\author{ $^{1,2}$F. A. Brito}
\email{fabrito@df.ufcg.edu.br}
\author{$^{3}$C. Wotzasek}
\email{clovis@if.ufrj.br}
\author{$^{3}$C. A. D. Zarro}
\email{carlos.zarro@if.ufrj.br}

\affiliation{$^{1}$Departamento de F\'{\i}sica, Universidade Federal de Campina Grande,\\
Caixa Postal 10071, 58429-900, Campina Grande, Para\'{\i}ba, Brazil.}
\affiliation{$^{2}$Departamento de F\' isica, Universidade Federal da Para\' iba,\\  Caixa Postal 5008, Jo\~ ao Pessoa, Para\' iba, Brazil.}
\affiliation{$^{3}$Instituto de F\' isica, Universidade Federal do Rio de Janeiro,\\  Caixa Postal 21945, Rio de Janeiro, 
Rio de Janeiro, Brazil.}
\affiliation{$^{4}$Grupo de F\' isica Te\'orica e Matem\'atica F\'{i}sica,  Departamento de F\'isica, Universidade Federal Rural do Rio de Janeiro, 23890-971, Serop\'edica, Rio de Janeiro, Brazil.}
\affiliation{$^{5}$Departamento de F\'{i}sica, Universidade Federal de Juiz de Fora, 36036-330, Juiz de Fora, Minas Gerais, Brazil}


\begin{abstract}
In this work we have used a Ho\v rava-Lifshitz scaling to rewrite a Lorentz-violating higher-order derivative electrodynamics controlled by a background four-vector $n_{\mu}$. The photon propagator was obtained and we have analyzed the dispersion relation and the observational results of gamma-ray burst (GRB) experiments were used. The limits of the critical exponent were discussed in the light of the GRB data and the physical implications were compared with the current GRB-Lorentz-invariance-violation literature. We show that the bound for the Lorentz-violating coupling for dimension-six operators, obtained from a Ho\v rava-Lifshitz scaling, is eight orders of magnitude better than the result found without considering a Ho\v rava-Lifshitz scaling, also this bound is nearby one, which is expected to be relevant phenomenologically.
\end{abstract}
\pacs{XX.XX, YY.YY} \maketitle


\section{Introduction}

Lorentz symmetry is one of the cornerstones of the Standard Model which describes very precisely the properties of elementary particles and interactions at energy scales below TeV, the energy scale attained by the Large Hadron Collider. However, the idea that it can be violated at high energies, presumably at Planck energy scale, $M_{\rm  P}\approx 1.22\times10^{19}$ GeV/c$^{2}$, has been investigated in many areas of theoretical physics (see, for example, Ref. \cite{Liberati} and references therein). In this manuscript, the Lorentz invariance violation (LIV) is introduced through high-order derivative operators which explicitly break Lorentz symmetry. These LIV terms are effective at the ultraviolet (UV) regime and they lead to modifications of dispersion relations for particles at high energies \cite{PospelovI} (see also Ref. \cite{PospelovIa}).

The study of higher order derivatives in the scenario of the LIV effective theory was initially proposed by Myers-Pospelov by using mass operators of dimension-five along 
with a nondynamical four-vector $n_\mu$ interacting with scalars, fermions, and photons fields \cite{PospelovI}. For the electromagnetic sector, their proposal reads:
\bea\label{eqMP}
S_{MP} = -\frac{\xi}{M_{\rm P}} \int d^{4} x\;  n^{\alpha}F_{\alpha\delta}n^{\gamma}\pa_{\gamma}n_{\beta}\tilde{F}^{\beta\delta},
\eea

\noindent $\xi$ is a dimensionless parameter, $\alpha,\beta,\gamma,\delta=0,1,2,3$, $F_{\alpha\beta}=\pa_{\alpha}A_{\beta}-\pa_{\beta}A_{\alpha}$ and $\tilde{F}^{\beta\delta}=\frac{1}{2}\varepsilon^{\beta\delta\rho\sigma}F_{\rho\sigma}$. One chooses the background four-vector $n_{\mu} $ as $n_{\mu}= (n_{0}\equiv1,\vec{0})$ which leads to spatial anisotropy in space-time. Then, one integrates by parts Eq.(\ref{eqMP}) and uses the Bianchi identities, $\pa_{[\alpha}F_{\beta\gamma]}=0, $ to obtain:
\bea\label{eq0a}
S_{MP} = -\frac{\xi}{2M_{\rm P}} \int d^{4} x\; \varepsilon^{ijk} A_{i} \pa_{t}^{2} F_{jk},\;\;\;\;\;\varepsilon^{ijk}\equiv \varepsilon^{0ijk}
\eea
where latin indices indicates the spatial components, $1,2,3$ and $F_{ij}$ is the spatial component of the electromagnetic field strength tensor.


As it can be inferred from Eq.(\ref{eq0a}) the Lorentz Invariance is broken around the scale $M_{\rm P}$, which may present additional theoretical as well as observational
challenges  to  such  models. Moreover, this theory can present problems associated with a non-renormalizable Lorentz-violating operator. Hence, in order to deal with this problem, one relies on a very anisotropic scale between space and time, which is also known as Ho\v{r}ava-Lifshitz field theory \cite{Horava:2008jf,Visser:2009fg}. In this approach, there is an explicit asymmetry between time and space coordinates so the Lorentz symmetry is broken. This asymmetry is controlled by a critical exponent $z$, which is  also dubbed as Lifshitz critical exponent. 

{The Ho\v{r}ava-Lifshitz theory has also been proposed as an alternative for a UV-completion of general relativity 
\cite{Horava:2009uw}. It is known that general relativity is not a UV-renormalizable interaction as its coupling constant has the following dimensions, $[G]=[M]^{-2}$, where $[M]$ denotes the mass unit. The main claim is that one can introduce an asymmetry between space and time, in this case, the dimension of time is  $[t]=[L]^{z}$ and the dimension of spatial coordinates is $[\vec{r}]=[L]$, where $[L]$ denotes the length unit. The effect of this asymmetry is the possibility of turning the coupling constant dimensionless after choosing judiciously the value of the critical exponent $z$. This interaction, now, is expected to be power countable UV-renormalizable. Indeed, in Ref.~\cite{Horava:2009uw}, it was shown that for $z\geq 3$ is sufficient to warrant the UV-renormalizability of general relativity. 
}
For $z=1$, one recovers the usual Lorentz symmetry for the low-energy limit.  Above the energy scale  $\Lambda_{HL}$ one expects to find LIV terms and $z\neq1$. In Ref.\cite{PospelovIIa}, it is argued that $\Lambda_{\rm HL}\ll M_{\rm P}$, which enhances the viability of observing experimentally LIV.

Our aim is to discuss the effects of LIV that can phenomenologically emerge below some intermediate scale, $\Lambda_{\rm HL}\sim 10^{10} {\rm GeV}$,
given by the process of separation between the Lifshitz-UV behavior and the effects of large energy scale (a discussion on the choice of scale for $\Lambda_{\rm HL}$ is found, for example, at Ref.\cite{PospelovIIa}).

In this paper, we take advantage of this procedure to rewrite the higher-derivative effective action (\ref{eq0a}) as a function
of the Lifshitz critical exponent $z$. This aims to find new phenomenological constraints on LIV (in the Lifshitz-UV behavior) by using recent measures of the gamma-ray busts (GRB) polarization. Note that previous studies (to large energy scales) apply certain corrections in the photons dispersion relations as a method to obtain constraints on LIV from polarization measurement of GRB \cite{grbGlaise, grbBoggs, grbJacobson} (see specific discussion, e.g., \cite{Gotz, LIVGRB02, GotzI}). Our goal is to revisit such investigations by influence of Lifshitz-UV behavior.

This paper is organized as follows: in section \ref{sec01} we have introduced the model and computed the photon propagator.  In section \ref{section-3} we have obtained the dispersion relation and used the results obtained by the gamma-ray burst observations.  The physical consequences were discussed.
 In section \ref{conclu} we have depicted the conclusions.


\section{The Gauge-Invariant Model}
\label{sec01}

Following the Ref.\cite{PospelovIIa}, a Horava-Lifshitz like modification of the Maxwell sector can be written as
\bea\label{eqSMHL}
S_{\rm M,HL} = -\frac{1}{2}\int dtd^{3}\vec{x}\;\Big[F_{0i}F^{0i}+\frac{1}{2}F_{ij}(-\Delta)^{z-1})F^{ij}\Big],
\eea

\noindent where $\Delta =  - \pa_{i}\pa^{i}= \vec{\pa}\cdot \vec{\pa}$, the metric used is $\text{diag }(1,-1,-1,-1)$ and $z$ is the Lifshitz critical exponent. For the Ho\v{r}ava-Lifshitz theory, in lenght units, one has $[t]=[L]^{z}$, $[\vec{x}]=[L]$, $[\pa_{t}]=[L]^{-z}$ and $[\pa_{i}]=[L]^{-1}$. As the action is dimensionless, one finds that $[A_{0}]=[L]^{-\frac{1}{2}(z+1)}$ and  $[A_{i}]=[L]^{\frac{1}{2}(z-3)}$. For $z=1$ one recovers the usual action for the free Maxwell field, as expected.  

To complete this model, a Ho\v{r}ava-Lifshitz like version of the Myers-Pospelov LIV action, Eq.(\ref{eq0a}), is introduced:
\bea
\label{eq:MPHL}
S_{\rm MP,HL}=-\frac{\xi}{2M_{\rm P}} \int dt d^{3} \vec{x}\; \varepsilon^{ijk} A_{i} \pa_{t}^{2}(\sqrt{-\Delta})^{z-1} F_{jk}
\eea
\noindent The Planck mass dimension is $[M_{\rm P}]=[L]^{-z}$. Following the Ref.\cite{Visser:2009fg}, one can rescale the dimensions as
\bea
\label{eq:reescale}
t\mapsto \Lambda_{\rm HL}^{-z+1}t;\;\;\pa_{t}\mapsto \Lambda_{\rm HL}^{z-1}\pa_{t};\;\;A_{0}\mapsto \Lambda_{\rm HL}^{\frac{1}{2}(z-1)}A_{0};\;\;A_{i}\mapsto \Lambda_{\rm HL}^{-\frac{1}{2}(z-1)}A_{i};\;\; M_P\mapsto \Lambda_{\rm HL}^{z-1}M_{\rm P}.
\eea

\noindent A Ho\v{r}ava-Lifshitz energy scale, $\Lambda_{\rm HL}$, which units are $[\Lambda_{\rm HL}]=[L]^{-1}$, is defined. After this procedure the quantities has the familiar dimensions: $[t]=[L]$, $[\pa_{t}]=[A_{0}]=[A_{i}]=[M_{\rm P}]=[L]^{-1}$.  Applying Eq. (\ref{eq:reescale}) to Eq.(\ref{eqSMHL}) one finds
\bea\label{eq:MaxwellHLreescaled}
S_{\rm M,HL}=-\frac{1}{2}  \int dt d^{3} \vec{x} \Big[F_{0i}F^{0i} + \frac{1}{2\Lambda_{\rm HL}^{2(z-1)}} F_{ij} (- \Delta)^{z-1}F^{ij}\Big],
\eea

\noindent which is exactly the model investigated in Ref.\cite{PospelovIIa}. After rescaling the variables, Eq.(\ref{eq:MPHL}) reads:
\bea
\label{ex01}
S_{\rm MP,HL}=-\frac{1}{2}  \int d^{4} x \Big[\frac{\xi  R_{\rm HLP}}{( \Lambda_{\rm HL})^{z}}  \varepsilon^{ijk} A_{k} \pa_{t}^{2}(\sqrt{-\Delta})^{z-1} F_{ij}\Big],
\eea

\noindent where $R_{\rm HLP}=\frac{\Lambda_{\rm HL}}{M_{\rm P}}$  is the ratio between the Ho\v{r}ava-Lifshitz cross-over scale $\Lambda_{\rm HL}$ and the Planck scale $M_{\rm P}$. This ratio can be considered very small ($R_{\rm HLP}\sim 10^{-9}$), providing a novel way of explicitly breaking supersymmetry without reintroducing fine-tuning \cite{PospelovIIa} (see also, Ref.\cite{PospelovII}).  

Now, the construction of the model used can be finished. The action of our model which has both Ho\v{r}ava-Lifshitz scaling and LIV terms is
\bea
\label{ex01}
S=-\frac{1}{2}  \int d^{4} x \Big[F_{0i}F^{0i} + \frac{1}{2\Lambda_{\rm HL}^{2(z-1)}} F_{ij} (- \Delta)^{z-1}F^{ij} +\frac{\xi  R_{\rm HLP}}{( \Lambda_{\rm HL})^{z}}  \varepsilon^{ijk} A_{k} \pa_{t}^{2}(\sqrt{-\Delta})^{z-1} F_{ij}\Big]
\eea

\noindent   Notice that at $\xi \to 0$, we recover a Lifshitz-type gauge invariant electrodynamics \cite{PospelovIIa}.

\subsection{Proof of gauge invariance of our model}

The gauge invariance of our model, Eq.(\ref{ex01}), can be examined similarly to Arnowitt-Deser-Misner formalism in Lifshitz gravity \cite{PospelovIIa,Desser}. We consider the following 
 decomposition of the fields: $A_{0}$ and $A_{i} = A^{\rm T}_{i} + \pa_{i}\varphi$ to rewrite the action (\ref{ex01}) as,

\bea\label{ex02}
S&=&\frac{1}{2}  \int d^{4} x \Biglb\{ A_{k}^{\rm T} \Big[\big(\pa_{t}^{2} + \Lambda_{\rm HL}^{- 2(z-1)} (- \Delta)^{z}\big)\eta^{ik} + 2 \xi  R_{\rm HLP}(\Lambda_{\rm HL})^{-z} \varepsilon^{ijk} \pa_{t}^{2}(\sqrt{-\Delta})^{z-1}\pa_{j}\Big]  A^{\rm T}_{i}+
\nonumber\\&& (A_{0} + \dot\varphi )\Delta (A_{0} + \dot\varphi )\Bigrb\}
\eea

\ni which makes explicit the following gauge symmetry: $A_{0} \to A^{\prime}_{0}= A_{0} + \dot\omega, \;\;\;\; \varphi\to \varphi^{\prime}= \varphi - \omega$, such as 
the usual gauge symmetry: $A_{\mu} \to A^{\prime}_{\mu}=A_{\mu} + \pa_{\mu}\omega$. 

\subsection{The Photon Propagator}

In this point, we derive the photon propagator associated with the action (\ref{ex01}) after the inclusion 
of the following gauge fixing term:
\bea\label{ex031}
{\cal L}_{\text{GF}}= -\frac{1}{2} \bigglb[ (\pa_{0}A_{0})^{2} + \frac{1}{\Lambda_{\rm HL}^{2(z-1)}} (- \Delta)^{z-1} (\pa_{i}A^{i})^{2} \biggrb]\,\,.
\eea
Thus, we choose the gauge condition $A_{0}=0$ \cite{PospelovIIa} to rewrite the  Lagrangian as
\bea\label{ex032}
{\cal L} = \frac{1}{2} A^{i} \Big[\big(\pa_{t}^{2} + \Lambda_{\rm HL}^{- 2(z-1)} (- \Delta)^{z}\big)\eta_{ik} - 2 \xi  R_{\rm HLP}( \Lambda_{\rm HL})^{-z}  \pa_{t}^{2}(\sqrt{-\Delta})^{z-1} \varepsilon_{ijk}\pa^{j} \Big]A^{k}
\eea
which we can identify the non-covariant photon kinetic operator in the following momentum representation (in our notation $ \vec{k}^{2} = |\vec{k}|^{2}= k^{2}$):
\bea\label{ex33}
(\hat{{\Delta}}^{-1}_{F})_{ik} = \big(-\omega^{2} + \Lambda_{\rm HL}^{- 2(z-1)} k^{2z}\big)\eta_{ik} + 2i \xi  R_{\rm HLP}( \Lambda_{\rm HL})^{-z}  \omega^{2}k^{(z-1)}\varepsilon_{ijk}k^{j}.
\eea
\ni Notice that the Feynman propagator resulting from the inversion is
\bea\label{ex34}
(\hat{\Delta}_{F}(k))^{ik}=\frac{1}{\hat{D}(k)}\Big[\big(\omega^{2} -  \Lambda_{\rm HL}^{-2(z-1)} k^{2z}\big)\eta^{ik} +
 2\xi  R_{\rm HLP}( \Lambda_{\rm HL})^{-z}  \omega^{2}k^{(z-1)}i \varepsilon^{ijk} k_{j}\Big]
\eea
\ni where
\bea\label{ex35}
\hat{D}(k) = \big(\omega^{2} - \Lambda_{\rm HL}^{-2(z-1)}k^{2z}\big)^{2} -  4 (\xi  R_{\rm HLP})^{2} (\Lambda_{\rm HL})^{-2z}  \omega^{4}k^{2z}.
\eea
\ni It is interesting to realize that when $z=1$ we have that the Eqs.(\ref{ex34}) and (\ref{ex35}) just recover the time-like background propagator connected to electromagnetic Myers-Pospelov Lagrangian in the Coulomb gauge \cite{Urrutia01}. The unitarity of the theory has been studied from the implementation of a physical cut-off both for the electron-positron scattering \cite{Reyes02} and tree level \cite{Scatena}. 


\section{Dispersion Relation and Phenomenological Aspects}
\label{section-3}

The dispersion relation is obtained from $\hat{D}(k)=0$ in Eq.(\ref{ex35}):      
\bea\label{ex06}
\big(\omega^{2} - \Lambda_{\rm HL}^{- 2(z-1)} {k}^{2z}\big)^{2} -  4 (\xi  R_{\rm HLP})^{2} (\Lambda_{\rm HL})^{-2z}  \omega^{4}{k}^{2z}=0
\eea

\subsection{Propagations Modes}

In order to derive a set of constraints on LIV from the vacuum birefringence effects by using specific observations of gamma-ray bursts, one has to solve the above dispersion relation to obtain the frequency solutions
\bea\label{ex07}
\omega_{\lambda} = \frac{{k}^{z}}{\Lambda_{\rm HL}^{(z-1)}\sqrt{1 - 2 \lambda \xi  R_{\rm HLP}  (\Lambda_{\rm HL})^{- z} {k}^{z}}}
\eea
with the two polarizations $\lambda=\pm$. 
Notice that the solutions correctly reproduces the usual ones in the limit $\xi \to 0$ at $z=1$.
For $\xi \neq 0 $ at $z=1$, we have solutions associated with the effects produced by dimension-five operators \cite{PospelovI} (see also Ref. \cite{Reyes02}).

Now we will carry out an expansion of the Eq.(\ref{ex07}) from the use of the following condition: $k^{z} \ll1/ 2 \xi  R_{\rm HLP} ( \Lambda_{\rm HL})^{ -z}$. This leads to
\bea\label{ex08}
\omega_{\lambda} \approx \frac{{k}^{z}}{(\Lambda_{\rm HL})^{z-1}} +  \frac{\lambda(\xi R_{\rm HLP}) {k}^{2z}}{(\Lambda_{\rm HL})^{2z-1}},
\eea
\ni which can be considered approximated solutions associated with the following modified dispersion relation:
\bea\label{ex081}
\omega^{2} - \frac{{k}^{2z}}{(\Lambda_{\rm HL})^{2(z-1)}} -  \frac{2\lambda (\xi R_{\rm HLP})}{(\Lambda_{\rm HL})^{3z-2}} {k}^{3z}=0.
\eea
For $z=1$, we recover the cubic modifications reported in Ref.\cite{PospelovI}. And, for $z > 1$ we find new expressions due to the direct influence of Lifshitz critical exponent.

Notice that the solutions (\ref{ex08}) can be decomposed into two sectors: $\omega(k)\to \omega_{\text{nb}}(k) + \omega_{\text{b}_\pm} (k) $, a nonbirefringent $( \omega_{\text{nb}}(k))$ and another birefringent $(\omega_{\text{b}_\pm} (k))$. The nonbirefringent sector is given by
\bea\label{ex082}
\omega_{\text{nb}}(k) = - k  +  \frac{{k}^{z}}{(\Lambda_{\rm HL})^{z-1}}.
\eea
For $z=1$ such nonbirefringent quantity  disappears. However to  $z > 1$, the dispersion relation can be associated with a specific model which breaks the gauge invariance (see, e.g.,  \cite{PospelovIIa} ). 
The birefringent sector can be written in the form
\bea\label{ex081}
\omega_{b_\pm} (k) = k \pm \frac{(\xi R_{\rm HLP}) {k}^{2z}}{(\Lambda_{\rm HL})^{2z-1}},
\eea

\ni which leads to a rotation of the polarization during the propagation of linearly polarized photons. 
For Lifshitz critical exponent $z= 1$, we  found the correction of order $ \xi (M_{\rm P})^{-1} k^{2}$ connected to operators of dimension-five.  And, to $z = 2$ we found the correction of order $ (\xi R_{\rm HLP}) (\Lambda_{\rm HL})^{-3} k^{4}$ associated with operators of dimension-seven.  Moreover, if we consider an intermediate critical exponent $z=3/2$, we found the correction of order $ (\xi R_{\rm HLP}) (\Lambda_{\rm HL})^{- 2} k^{3}$ associated with operators of dimension-six.


\subsection{ Constraints on LIV and gamma-ray burst}

The vacuum birefringence is an important effect to obtain bounds on LIV parameters \cite{Liberati}. Thus, assuming the dispersion relation given in Eq.(\ref{ex081}),  one finds that the direction of polarization rotates during propagation along a distance $d$ and is given by 
\bea\label{th01}
\Delta\theta(d)=\frac{\big(\omega_{\text{b}_{+}}(k)- \omega_{\text{b}_{-}}(k)\big)d}{2} \approx \frac{\xi}{2} \frac{ (R_{\rm HLP}) {k}^{2z}\, d} {(\Lambda_{\rm HL})^{2z-1}}.
\eea 
The above expression relates the degree of radiation polarized with phenomenon of birefringence. Consequently, 
the parameter which control the LIV is given as
\bea\label{th011}
\xi \approx \frac{2  \Delta\theta}{d}\frac{(M_{\rm P})^{2z-1}}{ \big(k_{2}^{2z} - k_{1}^{2z}\big)}\big(R_{\rm HLP}\big)^{2(z -1)}.
\eea

Now we are able to impose the upper bound for the  $\xi -$LIV parameter at Lifshitz point. To this, we use the recent determinations of the distance of the gamma-ray burst GRB 041219A, for which a high degree of polarization is
observed in the prompt emission. More informations see the detections derived in real time by the {\it INTEGRAL} Burst Alert System (IBIS) \cite{GRB01, GRB02, GRB03}. 

The measurement values to be used are the following:  $\Delta\theta(d)=47^{\circ}$ to degree of polarization derived from the measures made along the burst duration and  $d= 85\,{\rm Mpc} = 2.6 \times 10^{26} {\rm cm}$ to lower limit luminosity distance (corresponding to a redshift of $z_{red}= 0.02$) \cite{Gotz, LIVGRB02}. We also consider, $L_{Pl}\approx 1.62 \times 10^{-33}{\rm cm}$  as the Planck length and $M_{\rm P}\approx 1.22 \times 10^{19}{\rm GeV}$ as the Planck energy so that as a convenience, we can rewrite the limit luminosity distance as $d= 85\,{\rm Mpc} = 1.61 \times 10^{59} (M_{\rm P})^{-1}$. Thus, we can determine an upper limit to the LIV effect as being
\bea\label{th031}
\xi < \frac{1.02 \times 10^{-59} M_{\rm P}^{2z}}{(k_{2}^{2z} - k_{1}^{2z})}\big(R_{\rm HLP}\big)^{2(z-1)}.
\eea
In this point, we assume the following values to the energies: $k_{1}\sim 100 {\rm KeV}\sim 1 \times 10^{-4}{\rm GeV}\sim 0.81 \times 10^{-23} M_{\rm P}$,  $k_{2}\sim 350 {\rm KeV}\sim 3.5 \times 10^{-4}{\rm GeV}\sim 2.9 \times 10^{-23} M_{\rm P}$. Thus, as we have assumed that $R_{\rm HLP}\sim 10^{-9}$,  Eq.(\ref{th031}) is written as:
\bea\label{th032}
\xi_{(z)} \lesssim \frac{1.02 \times 10^{28z - 41}}{\big((2.9)^{2z} - (0.81)^{2z}\big)}
\eea
which corresponds to an anisotropic upper-limit directly controlled by the Lifshitz critical exponent.

\begin{figure}[h!]
\hspace{-2.0cm}
\includegraphics[scale=0.8]{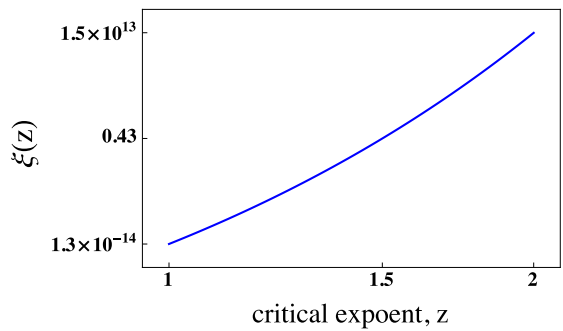} 
\vspace{-0.3cm}
\caption{The behavior of constraints on LIV for the following values: $z=1$, $z=3/2$ and $z=2$ respectively.}
\label{Fig1}
\end{figure}

To discuss the constraints brought by LIV we have plotted in Fig. \ref{Fig1}, the Eq.(\ref{th032}) as a function of the critical expoenent\footnote{Notice that $z$ is not constrained to be an integer \cite{Calcagni}.} $z$.
Notice that for $z=1$, a dimension-five operator, we find $\xi_{z=1} \lesssim 10^{-14}$, recovering the result obtained in \cite{LIVGRB02}. It is a very strict bound which may point to the non-existence of these kind of terms in a LIV action, or to the presence of some symmetry (maybe supersymmetry) acting to vanish this contribution of these operators. These conclusions were already discussed in Ref. \cite{LIVGRB02}. 

For a dimension-six operator, $z=\frac{3}{2}$, one finds that $\xi_{z=\frac{3}{2}} \lesssim 0{.}43$. If the Ho\v{r}ava-Lifshitz scaling is absent, the bound found in Ref.\cite{LIVGRB02} is $\xi \lesssim 2{.}61\times10^{8}$, this limit is too high to constrain any relevant $\xi$, which is expected to be nearby one. Notice, that not only the introduction of Ho\v{r}ava-Lifshitz anisotropic scaling improves the former result by eight orders of magnitude,  but also the bound found is almost of order one, where we hope to find a relevant LIV effect.
Notice that at  $z>\frac{3}{2}$, one finds that the limits found do not put any realistic restriction to the value of $\xi$ since, e.g., $\xi_{z=2} \lesssim 1{.}5\times 10^{13}$.

{Let us now discuss our phenomenological analysis by comparing with other studies of literature. We can consider a simple  generalization of the angle polarization given as \cite{LIVGRB02},
\bea\label{c01}
\Delta\theta(d) \approx \frac{\tilde\xi}{2} \frac{ {k}^{n}\, d} {M_{\rm Pl}^{n-1}},
\eea
where $n=2,3,...$, represents the order of operator. Now for the sake of comparison with the Eq. (\ref{th01}) we introduce the 
following relationship 
\bea\label{c02}
\frac{\xi}{\tilde\xi}= \frac{k^{n-2z}}{R_{\rm HLP}}\frac{(\Lambda_{\rm HL})^{2z-1}}{M_{\rm Pl}^{n-1}}.
\eea
Therefore, for operators of dimension-five $(n=2\; {\rm and}\; z=1)$ we find $\xi=\tilde\xi$, whereas
for operators of dimension-six $(n=3\; {\rm and}\; z=3/2)$ we find $\xi=(R_{\rm HLP})\tilde\xi$. This means that the Lifshitz scaling of LIV high derivative operator, i.e., $z > 1$, also rescales (by factor $10^{9}$) some phenomenological limits obtained by the usual model.

}

\subsection{Time delay of two photons }

Besides the phenomenological analyses considered above, at this point, we shall focus our attention on the time delay between two distinct photon flights. Such a delay has fundamental origin on the cosmological expansion itself and due to non-trivial dispersion relations. Since in our theory we have a modified dispersion relation it is interesting to address this issue as follows.

The dispersion relation (\ref{ex081}) leads to a modified speed of light for a photon with momentum $k$:
\bea\label{vg01}
v_{g}= \frac{d \omega_{b_{\pm}}}{d k}= 1 \pm 2 z (\xi R_{\rm HLP}) \Big(\frac{k}{\Lambda_{HL}}\Big)^{2z-1}. 
\eea 
Thus, the time delay between two photons with energy difference (for spatially flat Universe $\Omega_k=0$)  is given as
\bea\label{vg02}
\Delta t = \pm 2 z (\xi R_{\rm HLP}) ( \Lambda_{HL})^{1-2z} H_{0}^ {-1}\big(k_{2}^{2z-1} - k_{1}^{2z-1}\big)\int_{0}^{z_{red}} \frac{d z'}{\sqrt{\Omega_{\lambda}+
\Omega_{\rm m}(1 + z')^{3}}}  
\eea
where $H_0=70\, \rm {km}\, s^{-1}\rm{Mpc}^{-1}$ is the Hubble constant ($H_0^ {-1}=13.77\, \rm{Gyr}$). Now solving the integral for $\Omega_\lambda=0.7$, $\Omega_{\rm m}=0.3$ at a redshift $z_{red}=0.02$  and using the previously assumed values for $k_1$, $k_2$ and $R_{\rm HLP}$ we find
\bea
\Delta t \sim \pm \left(1.74\, z\,\xi_{(z)}\right)\times10^{-28z+21}\left((2.90)^{2z-1} - (0.81)^{2z-1}\right),\eea
where $\xi_{(z)}$ is defined in Eq.~(\ref{th032}). For $z=1$, $z=3/2$ and $z=2$,  we find respectively the time delay $\Delta t \lesssim 4.7\times 10^{-21}\,$s, $\Delta t \lesssim 8.7\times 10^{-21}\,$s and $\Delta t \lesssim 1.24\times 10^{-20}\,$s. In summary, these results show that contrary to the LIV parameter $\xi_{(z)}$, the time delay is not much sensitive with respect to the critical exponent $z$.
\section{Conclusions}
\label{conclu}

In this work, we have analyzed the LIV of a Lifshitz-scaling to a Maxwell-Myers-Pospelov model where the Lifshitz-type gauge invariant electrodynamics can be recovered.   The photon propagator was computed, the dispersion relations were derived and the frequency solutions for the two polarizations were obtained.  Also, 
a vacuum birefringence analysis was carried out.

Finally, we have obtained upper bounds for the $\xi$-LIV parameter for any Lifshitz critical exponent related to our model. These limits were determined using the observational results of the gamma-ray burst GRB 041219A.  For $z=1$, which corresponds to the usual dimensional-five operators, we found the same results already obtained in the literature. This presumably shows either that these type operators are absent in a LIV action, or there is a symmetry that vanishes their contributions in a LIV action. For $z=\frac{3}{2}$, which is related to dimension-six operators, it was shown that the bound for $\xi$ was improved in eight orders of magnitude from the same result where the Ho\v{r}ava-Lifshitz scaling is absent. Finally, we found that this bound is nearby one, which can be relevant from phenomenological point of view. For higher-dimensional operators, $z>\frac{3}{2}$, the bounds obtained do not really restrict any of the couplings $\xi_{z>\frac{3}{2}}$.


\begin{acknowledgments} 
 The authors thank CNPq (Conselho Nacional de Desenvolvimento Cient\' ifico e Tecnol\'ogico), Brazilian scientific support federal agency, for partial financial support and Grants numbers 302155/2015-5, 302156/2015-1 and 442369/2014-0.  E.M.C.A. thanks the kindness and hospitality of Theoretical Physics Department at Federal University of Rio de Janeiro (UFRJ), where part of this work was carried out.
\end{acknowledgments} 


\end{document}